\documentclass[amsmath,amssymb,aps,prd,showpacs,showkeys]{revtex4}%,preprint nofootinbib,
\usepackage{graphicx}% Include figure files
\usepackage{dcolumn}% Align table columns on decimal point
\usepackage{latexsym,color}
\begin{document}

\title{Theoretical and observational constraints \\ on the  mass-radius relations of neutron stars}

\author{Kuantay Boshkayev$^{1,2}$, Jorge A. Rueda$^{2}$ and Marco Muccino$^{2}$}
\affiliation{
$^1$IETP, Faculty of Physics and Technology, Al-Farabi Kazakh National University, Al-Farabi av. 71, 050040 Almaty, Kazakhstan\\ 
$^2$Dipartimento di Fisica and ICRA, Universit\`a di Roma "La Sapienza",  Piazzale Aldo Moro 5, I-00185 Roma, Italy}
\email{kuantay@mail.ru}

\date{\today}

\begin{abstract}

We investigate theoretical and observational constraints on the mass-radius relations for neutron stars. For that purpose we consider the model of neutron stars taking into considerations strong, weak, electromagnetic and gravitational interactions in the equation of state and integrate the structure equations within the Hartle-Thorne formalism for rotating configurations. On the basis of the theoretical restrictions imposed by general relativity, mass-shedding and axisymmetric secular instabilities we calculate the upper and lower bounds for the parameters of neutron stars. Our theoretical calculations have been compared and contrasted with the observational constraints and as a result we show that the observational constraints favor stiff equations of state.

\end{abstract}


\maketitle

\section{Introduction}

Neutron stars are very compact and  dense objects having average mass 1-2 $M_{\odot}$ (solar mass) and the average radius is around 10-15 km. The density in their center can exceed the nuclear density several times. They are an ideal laboratory which represents extreme conditions with high gravity, electromagnetic fields, density and pressure to test our theoretical models in nuclear and elementary particle physics \cite{shapirobook}. Probably, neutron stars are one of the fewest objects, where all fundamental interactions: strong, weak, electromagnetic and gravitational, take place \cite{shapirobook, haensel2007, potekhin2010}.

In this work, we consider the equilibrium structure of rotating neutron stars within the model proposed and recently extended by Belvedere et al. (2012, 2014) \cite{belvedere2012, belvedere2014} including the effects of rotation in terms of the Hartle-Thorne formalism \cite{hartle1967, har-tho:1968}. By fulfilling all the stability criteria and the latest observational and theoretical constraints on neutron star mass-radius relations, we computed the mass, radius, rotation frequency, angular momentum, quadrupole moment and other parameters of neutron stars.

Our paper is organized as follows: in Section \ref{sec:2}, we consider the external Hartle-Thorne solution and the neutron star models; in Section \ref{sec:3}, we discuss about the theoretical constraints on the mass-radius relations of neutron stars; in Section  \ref{sec:4}, we consider observational constraints. Finally, in Section \ref{sec:5}, we summarize our main results, discuss their significance, and draw our conclusions.

%%%%%%%%%%%%%%%%%%%%%%%%%%%%%%%%%%%%%%%%%%%%%%%%%%%%%%%%%%%%%%%%%%%%%%%%%%%%%%%%%%%%%%%%%%%%%%%%%%%%%%%%%%%%%%%%%%%%%%%%%%%%%%%%%%%%%%%%%%%%%%%%%%%%%%%%%%%%%%%%%%%%%%%%%%%%%%%%%%%%%%%%%%%%%%%%%%%%%%%%%%%%%%%%%%%%%%%%%%%%%%%%%%%%%%%%%
%
\section{The Hartle-Thorne metric and equation of state}\label{sec:2}

In the physics of compact objects the Hartle-Thorne solutions both internal and external are applied to study the main characteristics and calculate the basic parameters of rotating configurations starting from white dwarfs to quark stars \cite{haensel2007, boshkayev2013}. It allows one, for a given equation of state (EoS), to construct the mass-central density, the mass-radius relations and other relations in a simple way. Although it is an approximate solution of the Einstein field equations with accuracy up to the second order terms in the angular velocity of the star, it can be safely used to investigate the physical structure and properties of the relativistic objects in the strong field regime with intermediate rotation rate \cite{boshkayev2012, ber-etal:2005}.

The Hartle-Thorne metric \cite{har-tho:1968,boshkayev2012} describing the exterior field of a slowly rotating slightly deformed object is given by

\begin{eqnarray}\label{ht1}
ds^2&=&\left(1-\frac{2{ M }}{r}\right)\left[1+2k_1P_2(\cos\theta)+2\left(1-\frac{2{ M}}{r}\right)^{-1} \frac{J^{2}}{r^{4}}(2\cos^2\theta-1)\right]dt^2
\nonumber\\ && -\left(1-\frac{2{ M}}{r}\right)^{-1}\left[1-2\left(k_1-\frac{6 J^{2}}{r^4}\right)P_2(\cos\theta) -2\left(1-\frac{2{ M}}{r}\right)^{-1}\frac{J^{2}}{r^4}\right]dr^2
\nonumber\\ &&-r^2[1-2k_2P_2(\cos\theta)](d\theta^2+\sin^2\theta d\phi^2)+\frac{4J}{r}\sin^2\theta dt d\phi,
\end{eqnarray}
where
\begin{eqnarray}\label{ht2}
k_1&=&\frac{J^{2}}{{ M}r^3}\left(1+\frac{{ M}}{r}\right)+\frac{5}{8}\frac{Q-J^{2}/{ M}}{{ M}^3}Q_2^2\left(x\right)\ ,\nonumber \\
k_2&=&k_1+\frac{J^{2}}{r^4}+\frac{5}{4}\frac{Q-J^{2}/{ M}}{{ M}^2r}\left(1-\frac{2{ M}}{r}\right)^{-1/2}Q_2^1\left(x\right)\ ,\nonumber
\end{eqnarray}
and
\begin{eqnarray}\label{legfunc2}
Q_{2}^{1}(x)&=&(x^{2}-1)^{1/2}\left[\frac{3x}{2}\ln\frac{x+1}{x-1}-\frac{3x^{2}-2}{x^{2}-1}\right],\nonumber \\
Q_{2}^{2}(x)&=&(x^{2}-1)\left[\frac{3}{2}\ln\frac{x+1}{x-1}-\frac{3x^{3}-5x}{(x^{2}-1)^2}\right],
\end{eqnarray}
are the associated Legendre functions of the second kind, with $x=r/M -1$, and $P_2(\cos\theta)=(1/2)(3\cos^2\theta-1)$ is the Legendre polynomial. The constants ${M}$, ${J}$ and ${Q}$ are the total mass, angular momentum and  quadrupole moment of a rotating object, respectively.

The exterior Hartle-Thorne metric describes the gravitational field of any slowly and rigidly rotating, stationary and axially symmetric body. As one can see from Eq.~\ref{ht1} the exterior solution is given with accuracy up to the second order terms in the body's angular momentum, and first order in its quadrupole moment. Unlike other solutions of the Einstein field equations this solution possesses its internal counterpart. That is essential for the construction of the equilibrium configurations of rotating objects and calculate physical parameters inside and outside the sources of the gravitational fields.

There exist a number of models for neutron stars and correspondingly, the same number of equations of state. Depending on the nuclear compositions, theoretical assumptions and experimental data in nuclear physics the equations of state could be classified as soft, moderate and stiff. Different equations of state yield different mass-radius relations \cite{lattimer2001,lattimer2004,lattimer2007,lattimer2010,lattimer2016}. Hence there arises a natural question what EoS is more realistic? The only thing we know here is that the equation of state for neutron star must be constructed accounting for all fundamental interactions and the mass-radius relation must be in agreement with observational data. For this reason throughout this work we use the recent model of neutron stars formulated by Belvedere et al (2012) \cite{belvedere2012}.

\begin{figure}[htbp]
\centering
\includegraphics[width=0.75\columnwidth,clip]{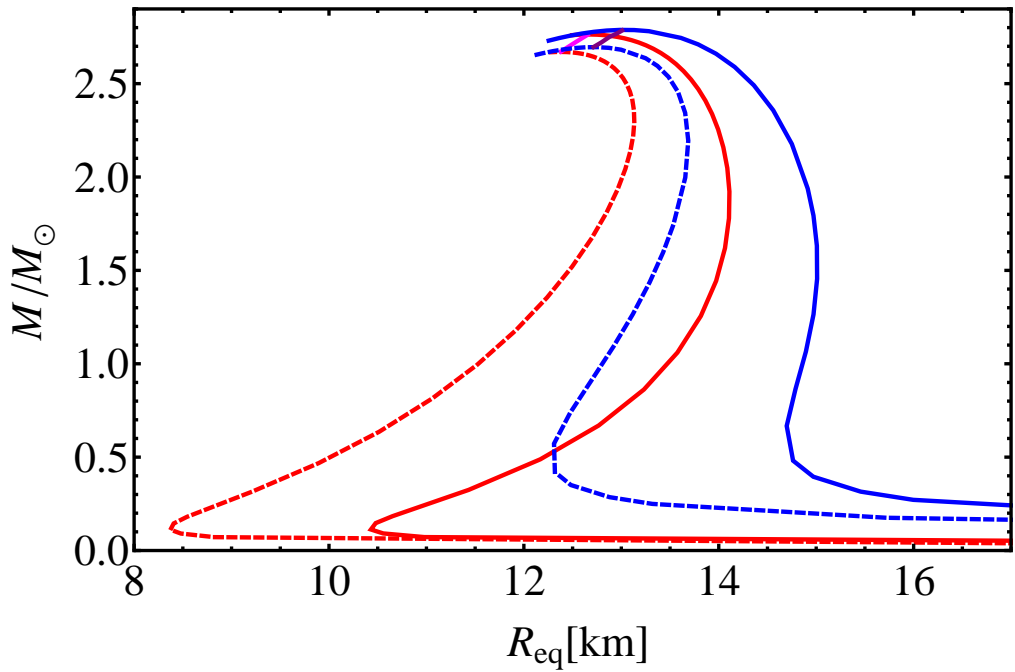}
\caption{Theoretical mass-radius relations presented in Belvedere et al. \cite{belvedere2014}. The red and blue  curves represent the configuration with global and local charge neutralities, respectively. Here dashed and solid curves are static and Keplerian sequences, respectively. The magenta and the purple lines represent the secular axisymmetric stability boundaries for the globally neutral and the locally neutral cases, respectively.} \label{fig:MR}
\end{figure}

By employing both interior and the exterior Hartle-Thorne solutions with the equations of state given in Ref. \cite{belvedere2012} we obtained the mass-radius relations for static and rotating configurations in both local and global charge neutrality cases. As one can see in Fig.~\ref{fig:MR} rotating neutron stars will possess larger mass and larger radius with respect to the static case.

% \cite{haensel2007, potekhin2010, yakovlev2016} 
%
\begin{figure}[htbp]
\centering
\includegraphics[width=0.75\columnwidth,clip]{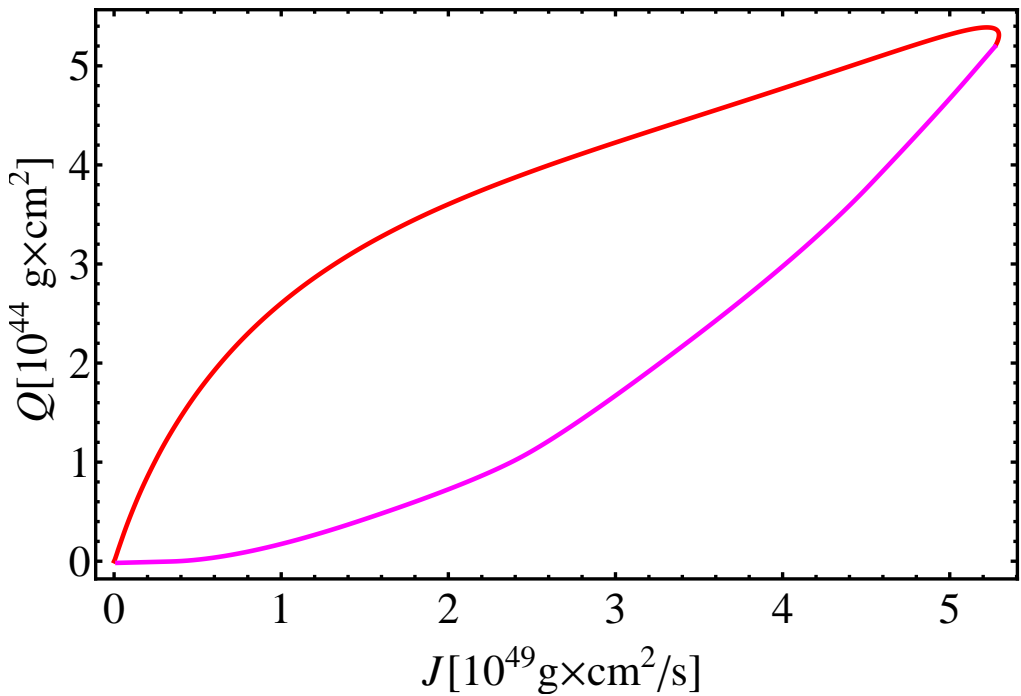}
\caption{The dependence of the quadrupole moment on the angular momentum of the rotating neutron star in the global charge neutrality cases. The red solid curve is the Keplerian sequences and the magenta curve is the secular axisymmetric stability boundary.} \label{fig:QJ}
\end{figure}

We also constructed the dependence of the quadrupole moment on the angular momentum in Fig.~\ref{fig:QJ}. Here we considered only global charge neutrality case, since for the local charge neutrality we have similar behavior. All possible values of $Q$ and $J$ for uniformly rotating neutron stars will be inside the loop. For vanishing angular velocity both $Q$ and $J$ will vanish simultaneously. By embedding in this diagram constant mass and constant frequency sequences one can infer either $Q$ or $J$ or both from observations.

While computing all these parameters we fulfilled stability criteria for rotating neutron star. Namely, the general relativistic instability related to the maximum mass, the mass-shedding limit (Keplerian limit) and the axisymmetric-secular instabilities have been taken into due account.

%%%%%%%%%%%%%%%%%%%%%%%%%%%%%%%%%%%%%%%%%%%%%%%%%%%%%%%%%%%%%%%%%%%%%%%%%%%%%%%%%%%%%%%%%%%%%%%%%%%%%%%%%%%%%%%%%%%%%%%%%%%%%%%%%%%%%%%%%%%%%%%%%%%%%%%%%%%%%%%%%%%%%%%%%%%%%%%%%%%%%%%%%%%%%%%%%%%%%%%%%%%%%%%%%%%%%%%%%%%%%%%%%%%%%%%%%
%
\section{Theoretical constraints}\label{sec:3}

In this section we discuss about theoretical constraints for neutron stars. First we consider the maximum mass.
The maximum possible mass of neutron star was calculated by Rhoades and Ruffini (1974) \cite{rhodes1974}. They assumed that general relativity is the correct theory of gravity and the Tolman-Oppenheimer-Volkoff equation determines the equilibrium structure, the equation of state is known below a fiducial value of the nuclear density, and that causality is not violated in the neutron star interior, namely that the speed of sound is subluminal at any density in the interior. As a result they obtained maximum 3.2 $M_{\odot}$ for unknown equation of state. Since then a lot attempts have been made to calculate maximum mass for different realistic equations of state \cite{rhodes1974, bethe1995, kalogera1996, heiselberg1999, schulze2006, gandolfi2012, chamel2013, bauswein2013,zhang2013, martinon2014, breu2016, maslov2016}. As expected, for realistic neutron stars the maximum mass is always less than 3.2 $M_{\odot}$.
%\cite{rhodes1974, bethe1995, kalogera1996, heiselberg1999, schulze2006, gandolfi2012, chamel2013, bauswein2013,zhang2013, martinon2014, breu2016, maslov2016}

\begin{figure}[htbp]
\centering
\includegraphics[width=0.75\columnwidth,clip]{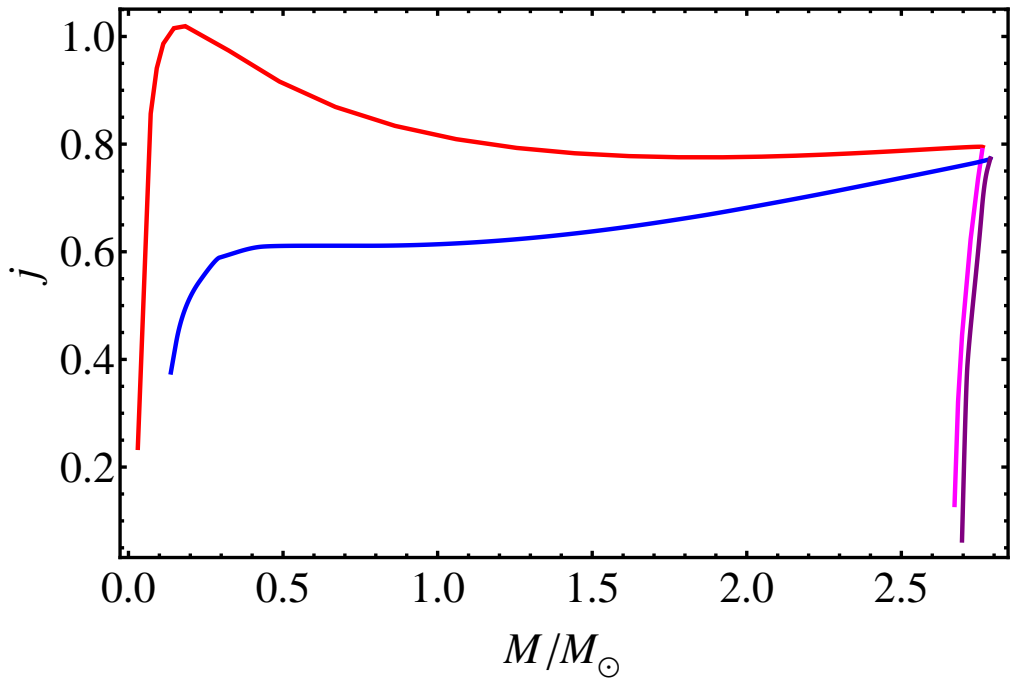}
\caption{Dimensionless angular momentum versus total mass. Red and blue solid curves are the Keplerian sequences, and magenta and purple curves are axisymmetric secular instability boundaries of both global and local neutrality cases, respectively. }\label{fig:jM_lgcn}
\end{figure}

For rotating neutron star the dimensionless angular momentum $j$ (spin parameter) can give an additional constraint. Relatively recently Lo \& Lin \cite{lolin2011} found that the maximum value of the dimensionless angular momentum $j_{max}$ of a neutron star uniformly rotating at the Keplerian sequence has an upper bound of about 0.7, which is essentially independent on the mass of neutron star as long as the mass is larger than about $1M_{\odot}$. However, the same parameter of a quark star does not have such a universal upper bound and could be larger than unity.

The dimensionless angular momentum has been also calculated by Cipolletta et al. (2015) \cite{cipolletta2015} for local charge neutrality cases with different equations of state and it has been also shown to be $j\approx0.7$ independent of the equation of state.

Furthermore, Qi et al. \cite{qi2014} extended the analyses of Lo \& Lin \cite{lolin2011} and Cipolletta et al. \cite{cipolletta2015} considering different kinds of uniformly rotating compact stars, including the traditional neutron stars, hyperonic neutron stars and hybrid stars. It was shown that the crust structure was a key factor to determine the properties of the spin parameter of the compact stars. When the crust EoSs are considered, $j_{max}\sim 0.7$ for $M > 0.5M_{\odot}$ is satisfied for three kinds of compact stars, no matter what the composition of the interior of the compact stars was.

When the crust EoSs are not included, the $j_{max}$ of the compact stars can be larger than $0.7$ but less than about $1$ for $M > 0.5M_{\odot}$. Consequently, according to Qi et al. \cite{qi2014} the crust structure provides the physical origin to the stability of $j_{max}$ but not the interior of the compact stars. The strange quark stars with a bare quark-matter surface are the unique one to have $j_{max} > 1$. Thus, one can identify the strange quark stars based on the measured $j > 1.0$, while measured $j\in(0.7, 1.0)$ could not be treated as a strong evidence of the existence of a strange quark star any more.

We also calculated the spin parameter using the model of neutron stars given by Belvedere et al. (2012). In Fig.~\ref{fig:jM_lgcn} the spin parameter is shown as a function of the total mass. Clearly, the value of $j$ is different from those of Lo \& Lin \cite{lolin2011} since we used different approach and different EoS. Despite this, the behavior of $j$ is more similar to those ones of Qi et al. \cite{qi2014} as we have crusts in both local and global neutrality cases. However, for the global charge neutrality the thickness of the crust is thiner than for the local charge neutrality and that is the reason for the spin parameter to be different in these cases. 

\begin{table}[ht]
\centering
\caption{Maximum mass and corresponding radius, maximum frequency and minimum period of globally and locally
neutral neutron stars.}
\begin{center}\label{table1}
\begin{tabular}{|c|c|c|}\hline
Physical parameters & Global neutrality & Local neutrality \\ \hline
$M_{max}^{J=0}/M_{\odot}$ & 2.67 & 2.70 \\ \hline
$R_{max}^{J=0}$ (km) & 12.38 & 12.71 \\ \hline
$M_{max}^{J\neq0}/M_{\odot}$ & 2.76 & 2.79  \\ \hline
$R_{max}^{J\neq0}$ (km) & 12.66 & 13.06 \\ \hline
$f_{max}$ (kHz) & 1.97 & 1.89 \\ \hline
$P_{min}$ (ms) & 0.51 & 0.53 \\ \hline
\end{tabular}
\end{center}
\end{table}

In Table~\ref{table1} we show upper bounds for static and rotating neutron stars within the model proposed by Belvedere et al (2012). Here we have stiff equation of state and correspondingly the maximum mass is larger than 2.6$M_{\odot}$ and smaller than 3.2$M_{\odot}$.

%%%%%%%%%%%%%%%%%%%%%%%%%%%%%%%%%%%%%%%%%%%%%%%%%%%%%%%%%%%%%%%%%%%%%%%%%%%%%%%%%%%%%%%%%%%%%%%%%%%%%%%%%%%%%%%%%%%%
%%%%%%%%%%%%%%%%%%%%%%%%%%%%%%%%%%%%%%%%%%%%%%%%%%%%%%%%%%%%%%%%%%%%%%%%%%%%%%%%%%%%%%%%%%%%%%%%%%%%%%%%%%%%%%%%%%%%
%
\section{Observational constraints}\label{sec:4}

According to observations, the most recent and stringent constraints to the mass-radius relation of neutron stars are provided from data for pulsars by the values of the largest mass, the largest radius, the highest rotational frequency, and the maximum surface gravity \cite{trumper2011}.

Up to now the largest neutron star mass measured with a high precision is the mass of the 39.12 millisecond pulsar PSR J0348+0432, $M=2.01 \pm 0.04 M_\odot$ \cite{antoniadis2013}. The largest radius is given by the lower limit to the radius of RX J1856-3754, as seen by an observer at infinity $R_\infty = R [1-2GM/(c^2 R)]^{-1/2} > 16.8$ km \cite{trumper2004}; it gives the constraint $2G M/c^2 >R-R^3/(R^{\rm min}_\infty)^2$, where $R^{\rm min}_\infty=16.8$ km. The maximum surface gravity is obtained by assuming a neutron star of $M=1.4M_\odot$ to fit the Chandra data of the low-mass X-ray binary X7, it turns out that the radius of the star satisfies $R=14.5^{+1.8}_{-1.6}$ km, at 90$\%$ confidence level, corresponding to $R_\infty = [15.64,18.86]$ km, respectively \cite{heinke2006}. The maximum rotation rate of a neutron star has been found to be $\nu_{\rm max} = 1045 (M/M_\odot)^{1/2}(10\,{\rm km}/R)^{3/2}$ Hz \cite{lattimer2004}. The fastest observed pulsar is PSR J1748-2246ad with a rotation frequency of 716 Hz \cite{hessels2006}, which results in the constraint $M \geq 0.47 (R/10\,{\rm km})^3 M_\odot$.
\begin{figure}[htbp]
\centering
\includegraphics[width=0.75\columnwidth,clip]{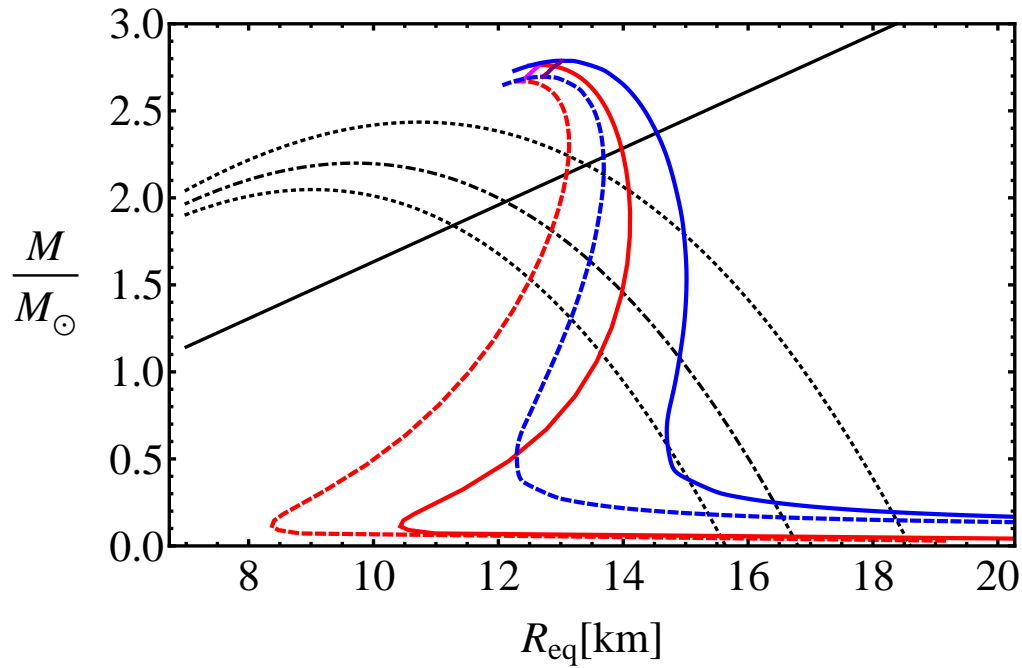}
\caption{Observational constraints on the mass-radius relation given by Tr{\"u}mper \cite{trumper2011} and the theoretical mass-radius relation presented in Fig \ref{fig:MR}. The solid black curve is the observed upper limit of the surface gravity, the  dotted-dashed black curve corresponds to the lower limit to the observed radius, and the dotted curves are the 90$\%$ confidence level contours of constant $R_\infty$.} \label{fig:MR_constr}
\end{figure}

From a technical or practical standpoint, in order to include the above observational constraints in the mass-radius diagram it is convenient to rewrite them for a given range of the radius (for instance, 6 km $\leq$ R $\leq$ 22 km) as follows: \\ 1. The maximum mass:
\begin{equation}\label{maxmass}
\frac{M}{M_{\odot}}=2.01.
\end{equation}
2. The maximum surface gravity:
\begin{equation}
\frac{M}{M_{\odot}}<2.4\times 10^5\frac{c^2}{G}\frac{R}{M_{\odot}}.
\end{equation}
3. The lower limit for the radius surface gravity:
\begin{equation}
\frac{M}{M_{\odot}}=\frac{10^5}{2}\frac{c^2}{G}\frac{R}{M_{\odot}}\left(1-\frac{R^2}{(R^{\rm min}_\infty)^2}\right).
\end{equation}
4. The maximum rotation rate:
\begin{equation}\label{freqmax}
\frac{M}{M_{\odot}}>\frac{0.47}{10^3}R^3.
\end{equation}
Note, that the last formula is valid only for the static mass-radius relations, since $R$ is the static radius. In order to include this constraint in the rotating mass-radius relation one should construct a constant frequency sequence for the fastest spinning pulsar with 716 Hz. For the sake of generality, we can just require that equilibrium models are bound by the Keplerian sequence (see Refs.~\cite{belvedere2014, cipolletta2015} for details). In all expressions above (\ref{maxmass}-\ref{freqmax}) the mass is normalized with respect to the solar mass $M_{\odot}$ and the radius is expressed in km.

In Fig.~\ref{fig:MR_constr} we superposed the observational constraints introduced by Tr{\"u}mper~\cite{trumper2011} with the theoretical mass-radius relations presented here and in Belvedere et al. \cite{belvedere2012, belvedere2014} for static and uniformly rotating neutron stars. Any realistic mass-radius relation should pass through the area delimited by the solid black, the dotted-dashed black, the dotted curves and the Keplerian sequences. From here one can clearly see that the above observational constraints show a preference on stiff EoS that provide largest maximum masses for neutron stars. From the above constraints one can infer that the radius of a canonical neutron star of mass $M = 1.4M_{\odot}$ is strongly constrained to $R\geq12$ km, disfavoring at the same time strange quark matter stars. It is evident from Fig.~\ref{fig:MR_constr} that mass-radius relations for both the static and the rotating case presented here, are consistent with all the observational constraints.

\section{Conclusion}\label{sec:5}
In this work we have considered the local and global neutrality cases in the model of neutron stars formulated by Belvedere et al. (2012). We also constructed the mass-radius diagram for rotating neutron stars on the basis of the work of Belvedere et al. (2014) within the Hartle-Thorne formalism. In addition, we calculated the maximum rotating mass, corresponding radius, minimum rotation period, dimensionless angular momentum, quadrupole moment and other crucial parameters of rotating neutron stars.

Furthermore, we considered theoretical constraints in the literature imposed on the mass-radius relations. Namely, we discussed about the maximum possible mass and maximum masses depending of the model of neutron stars, minimum periods, maximum dimensionless angular momentum, the relation between angular momentum and quadrupole moment etc. All these parameters are model dependent. Equations of state based on different models give different maximum and minimum values for all parameters.

In order to favor or disfavor some models we considered observational constraints on the mass-radius relations of neutron stars related to the maximum observed mass, maximum surface gravity, largest mass, maximum rotation frequency. All these constraints are important not only in the physics of neutron stars, but also in nuclear physics to test theoretical hypothesis and assumptions made in the construction of the equations of state. As a result all observations favor stiff equations of state as indicated by Yakovlev (2016) \cite{yakovlev2016}.

The results of this work can be applied to the investigation of the X-ray phenomena occurring in the accretion disks around neutron stars such as quasi periodic oscillations \cite{pachon2012}. Combining both the quasi periodic oscillations data from low X-ray binary systems and physics of compact objects one can extract information not only on the properties of the accretion disks, but also infer the parameters of neutron stars and constrain the equations of state \cite{boshkayevqpos2014, boshkayevqpos2015, pappas2015, stulik2015, stulik2016, boshkayev2016, chen2016}.

Finally, the correct determination of neutron star critical mass, including its crust, plays also a very important role in understanding the progenitors of long gamma-ray burst (GRB), proposed to originate in binary systems composed of an evolved star exploding as a Ib/c supernova and triggering a hypercritical accretion process onto a companion neutron star \cite{Fryer2014}, and short GRBs, originating from binary neutron star mergers. In both cases two outcomes are possible depending on whether or not the accretion process or the merger can push the neutron star or the merged core, respectively, beyond the critical mass \cite{2015ApJ...798...10R,2015ApJ...808..190R}.

\subsection*{Acknowledgements}

This work was supported by program No F.0679 of grant No 0073 and the grant for the university best teachers-2015 of the Ministry of Education and Science of the Republic of Kazakhstan. K.B. acknowledges ICRANet for hospitality.

\end{document}